\documentclass{article}
\usepackage{spconf,amsmath,graphicx,bm,cite}
\usepackage{booktabs}
\usepackage{braket}
\usepackage{breqn}

\usepackage[absolute,showboxes]{textpos}

\TPMargin{5pt}

\newcommand{\copyrightstatement}{
    \begin{textblock}{0.84}(0.08,0.93)    
         \noindent
         \footnotesize
         \copyright  2019 IEEE.  Personal use of this material is permitted.  Permission from IEEE must be obtained for all other uses, in any current or future media, including reprinting/republishing this material for advertising or promotional purposes, creating new collective works, for resale or redistribution to servers or lists, or reuse of any copyrighted component of this work in other works.
    \end{textblock}
}

\title{Unsupervised Training for Deep Speech Source Separation  with \\ 
Kullback-Leibler Divergence Based Probabilistic Loss Function
}
\name{Masahito Togami$^1$, Yoshiki Masuyama$^2$\thanks{This work was done while Yoshiki Masuyama and Yu Nakagome were  interns at LINE Corporation.}, Tatsuya Komatsu$^1$, and Yu Nakagome$^2$ 
}
\address{
  $^1$LINE Corporation, Tokyo, Japan\\
  $^2$Waseda University, Tokyo, Japan
  }
\begin{document}
\ninept
\maketitle
\begin{abstract}
In this paper, we propose a multi-channel speech source separation 
 with a deep neural network (DNN) which is trained 
 under the condition that no clean signal is available. As an alternative to a clean signal, 
 the proposed method adopts an estimated speech signal by an unsupervised speech source separation with a 
 statistical model. 
 As a statistical model of microphone input signal, we adopts  a time-varying spatial covariance matrix (SCM)  model  which includes 
 reverberation and background noise submodels so as to achieve robustness against reverberation and background noise. 
 The DNN infers intermediate variables which are needed for constructing the time-varying SCM.  
 Speech source separation is performed in a probabilistic manner 
  so as to avoid overfitting to separation error. 
Since there are multiple intermediate variables, a loss function which evaluates a single intermediate variable is not applicable. 
Instead,  the proposed method adopts a loss function which evaluates the output probabilistic signal directly based on Kullback-Leibler Divergence (KLD). 
Gradient of the loss function can be back-propagated into the DNN through all the intermediate variables.  
Experimental results under  reverberant conditions show that the proposed method can 
train the DNN efficiently  even when the number of training utterances is small, i.e., 1K.

\end{abstract}
\section{Introduction}
\label{sec:intro}
\copyrightstatement
Microphone input signal in teleconferencing systems, speech diarization systems, and automatic 
speech recognition systems is typically a mixture of multiple speech sources 
and it is also contaminated by room reverberation. Thus, speech source separation techniques have been highly spotlighted. 
As speech source separation techniques,  
blind source separation (BSS) \cite{duet, common,hiroe, kim,duong2010,SAWADA2013,kitamura,ito2016} has been 
actively studied.  Parameters which are needed for speech source 
separation can be optimized in an unsupervised manner with a statistical model. A speech source model is highly important 
for estimating a separation filter and solving the well-known inter-frequency permutation  problem \cite{sawadaperm}. 
There are two requirements for a speech source model in the BSS. At first, the speech source model should capture complicated spectral characteristics of a speech source.
Secondly, there should be a computationally efficient algorithm for optimizing parameters based on the speech source model. 
However, it is highly difficult to define a statistical model which fulfills these two requirements.

As supervised speech source separation techniques,  
recently, deep neural network (DNN) based approaches with a training dataset 
in which there are microphone input signal and corresponding oracle clean data 
have been widely studied, e.g., deep clustering (DC)  \cite{deepclustering2016,deepclustering2018}, 
permutation invariant training (PIT) \cite{PIT2017,yoshioka2018}, deep attractor network \cite{dan2017,dan2018}, and hybrid approaches with 
 BSS \cite{nugraha2016,nugraha_book,MOGAMI2018}. 
 DNN based approaches can capture complicated spectral characteristics of a speech source.
Parameter optimization can be done efficiently by forward calculation of the DNN.  
However, it is hard to obtain an oracle clean data in a target environment. 
Thus, it is highly required to train the DNN by utilizing only observed microphone input signals which contain multiple speech sources  
without an oracle clean data.



Recently, unsupervised DNN training techniques have been proposed \cite{lucas2019,tzinis2019}. 
These techniques estimate a time-frequency mask based on the DC.  
The DNN is trained without an oracle time-frequency mask.  
An estimated time-frequency mask by a BSS technique in an unsupervised manner is adopted as an alternative to the oracle time-frequency mask. 
In the BSS technique, a time-frequency mask is estimated under the assumption that  each component of the microphone input signal is sparse enough at the time-frequency domain.
However, when there are reverberation and background noise, the sparseness assumption does not hold  
and speech source separation performance degrades.

In this paper, we propose an unsupervised DNN training technique 
which utilizes an estimated speech signal by an unsupervised speech source separation with a time-varying spatial filter  
as an alternative to the clean speech signal. The time-varying spatial filter is constructed based  on a time-varying spatial covariance matrix (SCM)  model \cite{duong2010,togami_arxiv2019} 
which includes submodels about reverberation and background noise so as to increase speech source separation performance 
under reverberant and noisy environments. The proposed method also estimates a time-varying spatial filter  via the DNN. 
The DNN infers intermediate variables which are utilized for constructing the time-varying spatial filter. 
Since there are several errors in a separated signal, 
both the separated signal by the unsupervised method and the separated signal via the DNN are  modeled as a probabilistic signal 
  so as to avoid overfitting to the separation errors in traing phase.  The proposed method adopts a loss function which evaluates Kullback-Leibler Divergence (KLD)  between 
  the posterior probability density function (PDF) of the separated signal by the unsupervised method and 
  that of the separated signal via the DNN. 
Although there are multiple intermediate variables which should be inferred by the DNN,
gradient of the loss function can be back-propagated into the DNN through all the intermediate variables jointly,   
  thanks to evaluating  the output signal in the loss function.
Experimental results under  reverberant and noisy conditions show that the proposed method can 
train the DNN more effectively in an unsupervised manner than conventional methods 
even when the number of the training utterances is small, i.e., 1K.
 The proposed KLD loss function is also shown to achieve better performance than 
 the $l_{2}$ loss function that evaluates the output signal as a deterministic signal.

\section{Microphone input signal model}
In this paper, speech source separation is performed in a time-frequency domain.
Multi-channel microphone input signal 
$\bm{x}_{l,k}$ ($l$ is the frame index and $k$ is the frequency index) 
is modeled as follows:
\begin{equation}
\bm{x}_{l,k}=\sum_{i=1}^{N_{s}} \bm{c}_{i,l,k} + \bm{r}_{l,k}+ \bm{w}_{l,k},
\end{equation}
where  
$N_{s}$ is the number of the speech sources, $\bm{c}_{i,l,k}$ is the $i$th speech signal, 
$\bm{r}_{l,k}$ is the late reverberation term, and $\bm{w}_{l,k}$ is the multi-channel background noise term.
The objective of speech source separation is  estimation of $\bm{c}_{i,l,k}$. 
 
\section{Proposed method}

\subsection{Overview}
\begin{figure*}[t]
\vspace{-30pt}
 \begin{center}
 \includegraphics[width=13cm,clip]{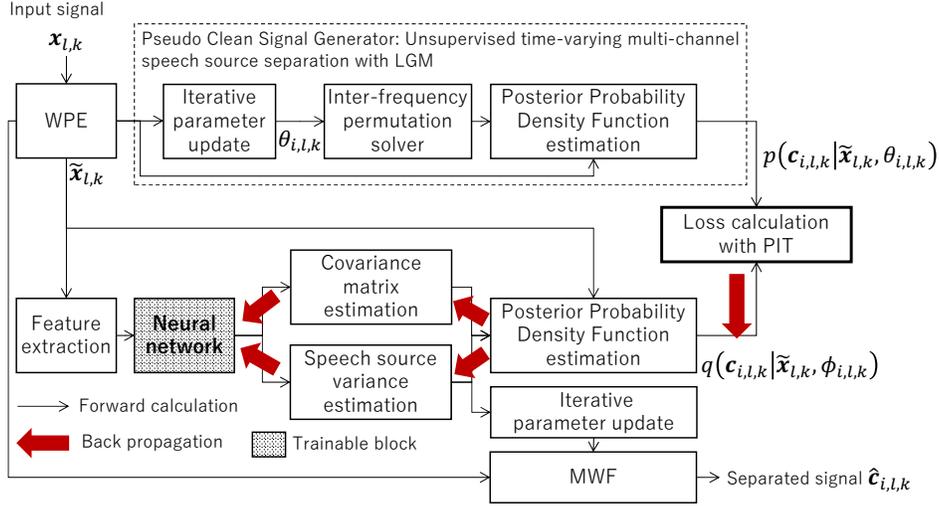}
 \end{center}
\vspace{-100pt}
 \caption{Block diagram of proposed method   \label{block_proposed}}

\end{figure*}

The proposed method trains a DNN which infers parameters 
of speech source separation without no clean data. 
Block diagram of the proposed method is shown in Fig.~\ref{block_proposed}.
The proposed method consists of two major parts. 
In each part, an input signal is a dereverberated signal by the Weighted Prediction Error (WPE) \cite{nakatani2010}. 
Let $\tilde{\bm{x}}_{l,k}=\bm{x}_{l,k}-\bm{W}_{k} \bm{X}_{l,k}$ be the output signal of the WPE, 
where $\bm{X}_{l,k}=[\begin{array}{ccc} \bm{x}_{l-D,k}^{T} & \cdots & \bm{x}_{l-L_{d}+1,k}^{T} \end{array}]^{T}$ 
($D$ is the tap-length of early reverberation and $L_{d}$ is the tap-length of the dereverberation filter $\bm{W}_{k}$).
The first part is a pseudo clean signal generator (PCSG).
As an alternative clean signal, 
the PCSG generates a separated speech signal  
in an unsupervised manner based on the local Gaussian modeling (LGM) \cite{duong2010}.
The PCSG regards the pseudo clean signal (PCS) as a probabilistic signal and estimates    
the posterior probability density function (PDF) of the PCS  
 $p(\bm{c}_{i,l,k} | \tilde{\bm{x}}_{l,k}, \theta_{k})$ in which  $\theta_{k}$ 
 is the separation parameter that is estimated in an iterative manner.
 The second part  is the DNN based estimation part of each speech source.
 In the DNN part,  each speech source is also regarded as a probabilistic signal and 
 the posterior PDF $q(\bm{c}_{i,l,k} | \tilde{\bm{x}}_{l,k}, \phi_{k})$
 is estimated, where $\phi_{k}$ is the separation parameter which is estimated via the DNN. 
 As the PCS and the estimated signal by the DNN are both probabilistic signals, 
 we evaluate the difference between the PCS and the estimated signal by a 
 loss function which evaluates a difference between two posterior PDFs.
 By consideration of uncertainty of the PCS and the estimated signal,
 gradient of the loss function propagates into the DNN not only through the mean vector but also 
 through the covariance matrix term of the posterior PDF inferred by the DNN, which leads to efficient DNN training.

\subsection{Pseudo Clean Signal Generator: Unsupervised speech source separation with local Gaussian modeling}
The LGM based speech source separation  \cite{duong2010} separates multiple speech sources 
assuming that the PDF of  each speech source belongs to a time-varying Gaussian distribution.
The PDF of the dereverberated signal is modeled as $
p(\tilde{\bm{x}}_{l,k})=\mathcal{N}(\tilde{\bm{x}}_{l,k}|\bm{0}, \bm{R}_{\tilde{x},l,k})$.
The multi-channel spatial covariance matrix (SCM) of the dereverberated signal $\bm{R}_{\tilde{x},l,k}$ is  modeled as follows:
\begin{equation}
\bm{R}_{\tilde{x},l,k}=\sum_{i} v_{i,l,k} \bm{R}_{i,k} + \bm{R}_{r, l,k} + \bm{R}_{n,k}, \label{cov_model}
\end{equation}
where the first term is the SCM of each speech source, 
$v_{i,l,k}$ is the time-frequency variance of the $i$th speech source, 
$\bm{R}_{i,k}$ is the multi-channel covariance matrix of the $i$th speech source, 
 the second term is the SCM of a residual late reverberation 
 which is not removed by the WPE, and the third term is the SCM of the background noise.
Reflecting that the amount of the late reverberation depends on 
the past speech source variance, the late reverberation term $\bm{R}_{r,l,k}$  is modeled as a convolution of
 the past time-varying 
speech source variance with the time-invariant covariance matrix \cite{togami_arxiv2019} as follows:
\begin{equation}
\bm{R}_{r, l,k}= \sum_{i,d=1}^{L_{r}} v_{i,l-d,k} \bm{H}_{i,d,k},
\end{equation}
where $L_{r}$ is the tap-length of the residual late reverberation and 
$\bm{H}_{i,d,k}$ is the time-invariant covariance matrix of the $i$th speech source.
The third term in Eq.~\ref{cov_model} is the time-invariant SCM of the background noise. 
Thus, $\theta_{k}$ is $\{\{v_{i,l,k} \}, \{\bm{R}_{i,k}\}, \{\bm{H}_{i,d,k}\}, \{\bm{R}_{n,k}\} \}$.
As all the PDFs are Gaussian distributions, the posterior PDF of the $i$th speech source is estimated 
as the following Gaussian distribution:
\begin{equation}
p(\bm{c}_{i,l,k}|\tilde{\bm{x}}_{l,k},\theta_{k})= \mathcal{N}(\bm{c}_{i,l,k}| \bm{\mu}_{p,i,l,k}, \bm{V}_{p,i,l,k}),
\end{equation}
where $\bm{\mu}_{p,i,l,k}$ and $\bm{V}_{p,i,l,k}$ are  calculated as $
 \bm{\mu}_{p,i,l,k}=\bm{W}_{i,l,k} \tilde{\bm{x}}_{l,k}, $ and $ 
\bm{V}_{p,i,l,k}=(\bm{I}-\bm{W}_{i,l,k}) v_{i,l,k} \bm{R}_{i,k},$   $\bm{I}$ is a $N_{m} \times N_{m}$  identity matrix ($N_{m}$ is  the number of the microphones), and 
$\bm{W}_{i,l,k}=v_{i,l,k} \bm{R}_{i,k} \bm{R}_{\tilde{x},l,k}^{-1}$ is the MWF.
The separation parameter $\theta_{k}$ is iteratively updated 
so as to maximize the log likelihood function $\sum_{l} \log p(\tilde{\bm{x}}_{l,k}|\theta_{k})$
 with  an  auxiliary 
function \cite{SAWADA2013,togami_arxiv2019}.
After $\theta_{k}$ update, the inter-frequency permutation problem is solved by \cite{MURATA20011}.

\subsection{Posterior PDF estimation via deep neural network}
\begin{figure}[htb]

 \begin{center}
 \includegraphics[width=6cm,clip]{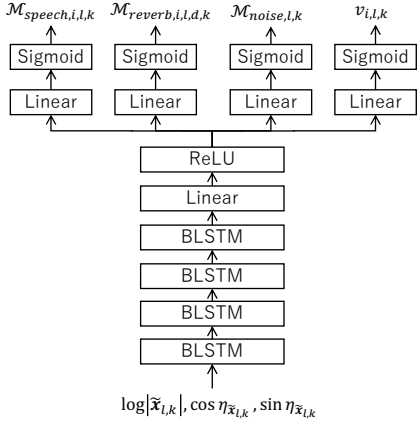}
 \end{center}
\vspace{-100pt}
 \caption{Deep neural network structure   \label{nn_proposed}}
\vspace{-10pt}
\end{figure}

In  the DNN part, 
the posterior PDF of each speech source is also estimated based on the LGM with the 
time-varying multi-channel SCM model 
defined in Eq.~\ref{cov_model}. The posterior PDF $q(\bm{c}_{i,l,k}|\tilde{\bm{x}}_{l,k},\phi_{k})= \mathcal{N}(\bm{c}_{i,l,k}| \bm{\mu}_{q,i,l,k}, \bm{V}_{q,i,l,k})$ 
is calculated with the estimated $\phi_{k}$. $\bm{\mu}_{q,i,l,k}$ and $\bm{V}_{q,i,l,k}$ are calculated in the same way as 
 $\bm{\mu}_{p,i,l,k}$ and $\bm{V}_{p,i,l,k}$, respectively.
 In the DNN part, the parameter $\phi_{k}$ is estimated via the DNN.
The DNN structure is shown in Fig.~\ref{nn_proposed}. The input feature 
is concatenation of log spectral of the dereverberated signal and phase difference between microphones $\eta_{\tilde{x},l,k}$.
Time-frequency masks and a time-frequency variance of each speech source are inferred via the DNN that contains 
four bidirectional long short term memory (BLSTM)  layers with 1200 hidden units 
and five dense layers. 
All of the covariance matrices are estimated via time-frequency masks inferred by the DNN, i.e., 
$\mathcal{M}_{speech,i,l,k}$, $\mathcal{M}_{reverb,i,l,d,k}$, and  $\mathcal{M}_{noise,l,k}$, 
as follows:
\begin{eqnarray}
\bm{R}_{i,k}&=&\frac{1}{\sum_{l} \mathcal{M}_{speech,i,l,k}} \sum_{l} \mathcal{M}_{speech,i,l,k} \tilde{\bm{X}}_{l,k}, \\
\bm{H}_{i,d,k}&=&\frac{1}{\sum_{l} \mathcal{M}_{reverb,i,l,d,k}} \sum_{l} \mathcal{M}_{reverb,i,l,d,k}\tilde{\bm{X}}_{l,k}, \\
\bm{R}_{n,k}&=&\frac{ 1}{\sum_{l} \mathcal{M}_{noise,l,k}} \sum_{l} \mathcal{M}_{noise,l,k}\tilde{\bm{X}}_{l,k}, 
\end{eqnarray}
where $\tilde{\bm{X}}_{l,k}=\tilde{\bm{x}}_{l,k} \tilde{\bm{x}}_{l,k}^{H}$ ($H$ is the Hermitian transpose of a matrix/vector). 

\subsubsection{Loss function for deep neural network training}
The loss function for the DNN training is set to a divergence between two posterior PDFs, 
i.e., the posterior PDF estimated by the LGM $p(\bm{c}_{i,l,k}|\tilde{\bm{x}}_{l,k},\theta_{k})$
 and the posterior PDF estimated via the DNN $q(\bm{c}_{i,l,k}|\tilde{\bm{x}}_{l,k},\phi_{k})$.
As a loss function, the proposed method adopts a Kullback-Leibler divergence 
$\mathcal{D}(p || q)$ defined as $
\mathcal{D}(p || q)=  \min_{f \in \Pi}  \sum_{i,l,k}  \mathcal{D}(p_{i,l,k} || q_{f(i),l,k}),$ where 
the utterance-level permutation invariant training (PIT) \cite{PIT2017} is utilized 
similarly to conventional supervised speech source separation \cite {yoshioka2018, MISD-M2019}, 
$\Pi$ is a set of possible permutations, and 
\begin{dmath}
\mathcal{D}(p_{i,l,k} || q_{j,l,k})= (\bm{\mu}_{q,j,l,k}-\bm{\mu}_{p,i,l,k})^{H}\bm{V}_{q,j,l,k}^{-1}(\bm{\mu}_{q,i,l,k}-\bm{\mu}_{p,i,l,k})
 +\text{tr}\Bigl(\bm{V}_{q,j,l,k}^{-1}\bm{V}_{p,i,l,k} \Bigr)+\log \frac{|\bm{V}_{q,j,l,k}|}{|\bm{V}_{p,i,l,k}|}-N_{m}.
 \label{loss}
\end{dmath}
It is shown that $\bm{V}_{q,j,l,k}$ acts as a regularization term in Eq.~\ref{loss}, 
 which leads to avoiding overfitting of the MAP estimate $\bm{\mu}_{q,i,l,k}$ to $\bm{\mu}_{p,i,l,k}$ that contains separation error, 
 and gradient of the loss function propagates not only through $\bm{\mu}_{q,i,l,k}$ but also through $\bm{V}_{q,j,l,k}$, which is favorable 
 for the DNN training of time-frequency masks.
 
\subsubsection{Output signal in inference phase}
In the inference phase, the parameter $\phi_{k}$ is inferred via a DNN.
After that, $\phi_{k}$ is iteratively updated so as to minimize the auxiliary function in the same way 
as the PCSG. 
Finally, the separated signal is obtained as 
a mean vector of the posterior PDF, $\hat{\bm{c}}_{i,l,k}=q(\bm{c}_{i,l,k}|\tilde{\bm{x}}_{l,k},\phi_{k})$, 
 by the MWF.

\section{Experiment}

\begin{table}[htb]
\vspace{-20pt}
\caption{Evaluation results of LGM based methods}
\label{evaluation_result_LGM}
\begin{center}
\begin{tabular}{ccccccc}\toprule
 $L_{r}$ & Loss  & SDR & SIR & CD & FWSeg.  & PESQ \\
  & Func.  & (dB) & (dB) & (dB) & SNR (dB)  & \\\midrule
\multicolumn{2}{c}{Unprocessed} & -2.01 & 0.52 & 5.60 & 	6.56  &  1.52  \\\midrule
$1$ & -  & 4.75 & 8.42 & 5.05  & \textbf{9.05}  & 1.90  \\
$4$  & -  & 4.12 & 8.19 & 5.12  & 8.20  & 1.82  \\
$8$  & -  & 3.84 & 7.76 & 5.17  & 7.93  & 1.78  \\\midrule
$1$  & $l_{2}$  & 5.14 & 8.99 & 5.05  & 8.79 & 1.92  \\
$4$  & $l_{2}$  & 4.87 & 8.71 & 5.04  & 8.14  & 1.89  \\
$8$ & $l_{2}$  & 3.62 & 6.02 & 5.27  & 7.04  & 1.75  \\\midrule
$1$  & KLD  & 5.44  & 9.74 & 4.95  & 8.63  & 1.98 \\
$4$  & KLD  & 5.53 & 9.84 & 4.89  & 8.88  & 1.99  \\
$8$  & KLD  & \textbf{5.71} & \textbf{10.27} & \textbf{4.86}  & 8.95  & \textbf{2.02}  \\\bottomrule 
\end{tabular}
\end{center}
\vspace{-30pt}
\end{table}
\begin{table*}[htb]
\caption{Comparison between CACGMM based methods and proposed method}
\label{evaluation_result_CACGMM}
\begin{center}
\begin{tabular}{ccccccccccc}\toprule
Approaches & Filtering & Loss Func. &  Phase Diff. & SDR (dB) & SIR (dB) & CD (dB) & FWSeg.SNR (dB)  & PESQ \\\midrule
\multicolumn{4}{c}{Unprocessed} & -2.01 & 0.52 & 5.60 & 	6.56  &  1.52 \\\midrule 
CACGMM &  Mask & - & - & 4.63 & 9.70 & 5.01  &  7.58  & 1.87 \\
CACGMM &  MVDR & - & - & 5.11 & 7.91 & 5.19  &  8.87  & 1.87 \\
CACGMM & Mask & DC & No & 3.03 & 5.26 & 5.37  & 6.46  & 1.65 \\
CACGMM & MVDR & DC & No & 3.35 & 4.36 & 5.42  & 7.57 & 1.68 \\
CACGMM & Mask & DC & Yes & 4.10 & 8.32 & 5.15  & 7.06  & 1.79 \\
CACGMM & MVDR & DC & Yes & 4.58 & 6.85 & 5.26  & 8.39  & 1.81 \\
CACGMM & Mask & MSA+PIT & No & 3.41 & 5.74 & 5.33  & 6.87  & 1.66 \\
CACGMM & MVDR & MSA+PIT & No & 3.63 & 4.92 & 5.41  & 7.61  & 1.70 \\
CACGMM & Mask & MSA+PIT & Yes & 4.64 & 8.87 & 5.04  & 7.76  & 1.85 \\
CACGMM & MVDR & MSA+PIT & Yes & 4.95 & 7.41 & 5.21  & 8.70  & 1.84 \\\midrule
LGM & MWF ($L_{r}=8$) & KLD & Yes & \textbf{5.71} & \textbf{10.27} & \textbf{4.86}  & \textbf{8.95}  & \textbf{2.02}  \\\bottomrule 

\end{tabular}
\vspace{-15pt}

\end{center}
\end{table*}

\subsection{Experimental setup}
Speech source separation performance of the proposed method was evaluated by using 
measured impulse responses in Multi-channel Impulse Response Database (MIRD) \cite{mird} and TIMIT speech corpus \cite{timit}.
In the training phase, TIMIT train corpus was utilized. 
In the evaluation phase, TIMIT test corpus was utilized. 
The reverberation time $RT_{60}$ was randomly set to 0.36 (sec) or 0.61 (sec). 
Sampling rate was set to 8000 Hz. 
The number of the microphone $N_{m}$ was set to $2$. The number of the speech sources $N_{s}$ was set to $2$. 
Two microphone indices were randomly selected for each sample both in the training 
phase and in the evaluation phase. 
A 3-3-3-8-3-3-3 spacing (cm) microphone array, a 4-4-4-8-4-4-4 spacing (cm) microphone array, and 
  a 8-8-8-8-8-8-8 spacing (cm) microphone array were utilized.
Frame size was 256 pt. Frame shift was 64 pt. The number of frequency bins was $129$.
The distance between speech sources and microphones was set to $1$ m. Azimuth of each talker 
is randomly selected for each utterance. The number of total training utterances was set to 1000, which is  
a smaller dataset than the conventional one, e.g., 30000 \cite{lucas2019}, because small number of required utterances is preferable 
 in practice. 
The number of total test utterances was 200. As a background noise signal, white Gaussian noise is added.
Signal to noise Ratio (S/N) was randomly set from 20 dB to 30 dB. 
S/N between two speakers was randomly set from -5 dB to 5 dB.
Mini-batch size was set to 128. Each utterance was split in every  100-frames segment. 
Neural network parameters were updated by $10000$ times. Adam optimizer \cite{Adam} (learning rate was $0.001$) with gradient 
clipping was utilized. The proposed method calculates  complex-valued gradient  by  
Tensorflow \cite{tensorflow2015-whitepaper}.
In each method, WPE was utilized (tap length was $16$ and  $D$ was set to $2$).
\subsection{Evaluation measures and comparative methods} 
We utilized Cepstrum distance (CD),  Frequency-weighted segmental 
SNR (FWSegSNR), and PESQ as dereverberation performance measures.
For speech source separation performance evaluation, 
we utilized SDR and SIR from BSS\_EVAL \cite{bsseval}.
Four methods were evaluated, i.e., 1) Conventional unsupervised training method with complex angular central Gaussian mixture model (CACGMM) \cite{lucas2019}: Time-frequency mask 
of each source is inferred with the sparseness assumption. This model does not have any reverberation model. A loss function which 
evaluates an intermediate variable is adopted. The DNN has four BLSTM layers. Only the output dense layer of the DNN is different from that of the proposed method. 
2) Unsupervised speech source separation based on LGM without DNN: The separation parameter is updated iteratively based on \cite{SAWADA2013}. 
This method is also identical to PCGS in the proposed method.
3) Unsupervised training  with LGM based PCGS and $l_{2}$ loss function: The loss function evaluates 
difference between the MAP estimate of the PCS posterior PDF and that of the estimated posterior PDF via the DNN, i.e., 
$L_{l_{2}}=\sum_{i,l,k} \lVert \bm{\mu}_{q,i,l,k}-\bm{\mu}_{p,i,l,k} \rVert^{2}$.
4) Proposed unsupervised training with LGM based PCGS and KLD loss function.

\subsection{Experimental results} 
At first, we evaluated three types of LGM based methods. 
The number of the covariance matrices of residual late reverberation 
$L_{r}$ was set to 1, 4, or 8. 
In Table \ref{evaluation_result_LGM}, 
experimental results for LGM based methods are shown.
The proposed unsupervised training methods with KLD loss function  
is shown to be more effective than the unsupervised training methods with $l_{2}$ loss function.
The proposed method also outperformed the  LGM without the DNN (PCGS). This result confirmed that 
the proposed method is robust against separation error of the PCGS. 
In the $l_{2}$ loss function cases, when $L_{r}$ is 4 or 8, performance was degraded.
It can be interpreted that the DNN parameters  were not correctly learned
 by back-propagation only through the mean vector of the posterior PDF. 
In the proposed KLD loss function cases, performance monotonically 
increased in accordance with the number of $L_{r}$. It is shown that 
back-propagation via the covariance matrix term of the posterior PDF is effective.

In Table \ref{evaluation_result_CACGMM},
we compared the proposed KLD loss function based method with $L_{r}=8$ 
and the conventional unsupervised DNN training method with CACGMM.  Unlike the proposed method,
 CACGMM does not have a reverberation model. 
Originally, a CACGMM based method without 
phase difference feature was proposed in \cite{lucas2019}. 
However, the proposed method utilizes phase difference between microphones 
as an input feature. To evaluate each method fairly, we also evaluated 
CACGMM based methods with the phase difference feature. 
In addition to deep clustering (DC) based methods in which the dimension of the embedding vector 
was set to $20$, PIT based methods which evaluate time-frequency masks were also evaluated. 
The time-frequency mask is evaluated by the magnitude spectrum approximation (MSA), because 
 the pseudo oracle time-frequency mask is real-valued and 
 the  phase-sensitive spectrum approximation (PSA) \cite{Erdogan2015} cannot be utilized. 
 We also evaluated the original CACGMM \cite{ito2016} without no DNN training. 
As an output signal, we evaluated time-frequency masking results and 
minimum variance distortionless response (MVDR) results. 
It is shown that the proposed method outperformed all variants of 
CACGMM based methods. This result confirmed that effectiveness of the proposed reverberation and background noise models 
 and DNN training with the proposed probabilistic loss function based on KLD. 

\section{Conclusions}
We proposed an unsupervised multi-channel speech source separation method in which 
the deep neural network (DNN) is trained with no oracle clean signal.
As a pseudo clean signal, the proposed method adopts the separated signal  
by the conventional unsupervised local Gaussian modeling. 
So as to reduce reverberation and background noise 
effectively, the proposed method estimates a time-varying 
covariance matrix of microphone input signal which contains 
reverberation and background noise components.  Since both the pseudo clean signal and the estimated signal via the DNN 
 are probabilistic signals, we proposed a loss function which evaluates the Kullback-Leibler divergence (KLD) 
between two posterior probability density functions.  
Experimental results showed that the proposed method can separate speech sources more accurately than the conventional methods
 under  a reverberant and noisy environment.

\bibliographystyle{IEEEbib}

\bibliography{mask,lgm,spbasic,gan,doa,kalman}

\begin{thebibliography}{10}

\bibitem{duet}
O.~Yilmaz and S.~Rickard,
\newblock ``Blind separation of speech mixtures via time-frequency masking,''
\newblock {\em IEEE Transactions on Signal Processing}, vol. 52, no. 7, pp.
  1830--1847, July 2004.

\bibitem{common}
P.~Common,
\newblock ``Independent component analysis, a new concept ?,''
\newblock {\em Signal Processing}, vol. 36, no. 3, pp. 287--314, April 1994.

\bibitem{hiroe}
A.~Hiroe,
\newblock ``Solution of permutation problem in frequency domain ica using
  multivariate probability density functions,''
\newblock in {\em Proceedings ICA}, Mar. 2006, pp. 601--608.

\bibitem{kim}
T.~Kim, H.T. Attias, S.-Y. Lee, and T.-W. Lee,
\newblock ``Independent vector analysis: an extension of ica to multivariate
  components,''
\newblock in {\em Proceedings ICA}, Mar. 2006, pp. 165--172.

\bibitem{duong2010}
N.Q.K. Duong, E.~Vincent, and R.~Gribonval,
\newblock ``Under-determined reverberant audio source separation using a
  full-rank spatial covariance model,''
\newblock {\em IEEE Trans. Audio Speech Lang. Process.}, vol. 18, no. 7, pp.
  1830--1840, 2010.

\bibitem{SAWADA2013}
H.~Sawada, H.~Kameoka, S.~Araki, and N.~Ueda,
\newblock ``Multichannel extensions of non-negative matrix factorization with
  complex-valued data,''
\newblock {\em IEEE Trans. Audio, Speech, and Language Process.}, vol. 21, no.
  5, pp. 971--982, May 2013.

\bibitem{kitamura}
D.~Kitamura, N.~Ono, H.~Sawada, H.~Kameoka, and H.~Saruwatari,
\newblock {\em Determined Blind Source separation with Independent Low-Rank
  Matrix Analysis}, chapter~6, pp. 125--155,
\newblock Springer Publishing Company, Incorporated, 2018.

\bibitem{ito2016}
N.~{Ito}, S.~{Araki}, and T.~{Nakatani},
\newblock ``Complex angular central gaussian mixture model for directional
  statistics in mask-based microphone array signal processing,''
\newblock in {\em EUSIPCO 2016}, Aug 2016, pp. 1153--1157.

\bibitem{sawadaperm}
H.~Sawada, S.~Araki, and S.~Makino,
\newblock ``Underdetermined convolutive blind source separation via frequency
  bin-wise clustering and permutation alignment,''
\newblock {\em IEEE Transactions on Audio, Speech, and Language Processing},
  vol. 19, no. 3, pp. 516--527, March 2011.

\bibitem{deepclustering2016}
J.R. Hershey, Z.~Chen, J.~Le Roux, and S.~Watanabe,
\newblock ``Deep clustering: Discriminative embeddings for segmentation and
  separation,''
\newblock in {\em ICASSP 2016}, 2016, pp. 31--35.

\bibitem{deepclustering2018}
Z.Q. Wang, J.~Le Roux, and J.R. Hershey,
\newblock ``Multi-channel deep clustering: Discriminative spectral and spatial
  embeddings for speaker-independent speech separation,''
\newblock in {\em ICASSP 2018}, 2018, pp. 1--5.

\bibitem{PIT2017}
D.~Yu, M.~Kolb{\ae}k, Z.~H. Tan, and J.~Jensen,
\newblock ``Permutation invariant training of deep models for
  speaker-independent multi-talker speech separation,''
\newblock in {\em ICASSP 2017}, March 2017, pp. 241--245.

\bibitem{yoshioka2018}
T.~{Yoshioka}, H.~{Erdogan}, Z.~{Chen}, and F.~{Alleva},
\newblock ``Multi-microphone neural speech separation for far-field
  multi-talker speech recognition,''
\newblock in {\em ICASSP 2018}, April 2018, pp. 5739--5743.

\bibitem{dan2017}
Z.~{Chen}, Y.~{Luo}, and N.~{Mesgarani},
\newblock ``Deep attractor network for single-microphone speaker separation,''
\newblock in {\em ICASSP 2017}, March 2017, pp. 246--250.

\bibitem{dan2018}
Y.~{Luo}, Z.~{Chen}, and N.~{Mesgarani},
\newblock ``Speaker-independent speech separation with deep attractor
  network,''
\newblock {\em IEEE/ACM Transactions on Audio, Speech, and Language
  Processing}, vol. 26, no. 4, pp. 787--796, April 2018.

\bibitem{nugraha2016}
A.A. Nugraha, A.~Liutkus, and E.~Vincent,
\newblock ``Multichannel audio source separation with deep neural networks,''
\newblock {\em IEEE/ACM Trans. Audio Speech Lang. Process.}, vol. 24, no. 9,
  pp. 1652--1664, 2016.

\bibitem{nugraha_book}
A.A. {Nugraha}, A.~{Liutkus}, and E.~{Vincent},
\newblock ``{Deep neural network based multichannel audio source separation},''
\newblock in {\em {Audio Source Separation}}. {Springer}, Mar. 2018.

\bibitem{MOGAMI2018}
S.~{Mogami}, H.~{Sumino}, D.~{Kitamura}, N.~{Takamune}, S.~{Takamichi},
  H.~{Saruwatari}, and N.~{Ono},
\newblock ``Independent deeply learned matrix analysis for multichannel audio
  source separation,''
\newblock in {\em EUSIPCO 2018}, Sep. 2018, pp. 1557--1561.

\bibitem{lucas2019}
L.~{Drude}, D.~{Hasenklever}, and R.~{Haeb-Umbach},
\newblock ``Unsupervised training of a deep clustering model for multichannel
  blind source separation,''
\newblock in {\em ICASSP 2019}, May 2019, pp. 695--699.

\bibitem{tzinis2019}
E.~{Tzinis}, S.~{Venkataramani}, and P.~{Smaragdis},
\newblock ``Unsupervised deep clustering for source separation: Direct learning
  from mixtures using spatial information,''
\newblock in {\em ICASSP 2019}, May 2019, pp. 81--85.

\bibitem{togami_arxiv2019}
M.~{Togami},
\newblock ``Multi-channel time-varying covariance matrix model for late
  reverberation reduction,''
\newblock {\em arXiv:1910.08710}, 2019.

\bibitem{nakatani2010}
T.~Nakatani, T.~Yoshioka, K.~Kinoshita, M.~Miyoshi, and B.~H. Juang,
\newblock ``Speech dereverberation based on variance-normalized delayed linear
  prediction,''
\newblock {\em IEEE Transactions on Audio, Speech, and Language Processing},
  vol. 18, no. 7, pp. 1717--1731, Sept 2010.

\bibitem{MURATA20011}
N.~{Murata}, S.~{Ikeda}, and A.~{Ziehe},
\newblock ``An approach to blind source separation based on temporal structure
  of speech signals,''
\newblock {\em Neurocomputing}, vol. 41, no. 1, pp. 1 -- 24, 2001.

\bibitem{MISD-M2019}
M.~{Togami},
\newblock ``Multi-channel {Itakura} {Saito} distance minimization with deep
  neural network,''
\newblock in {\em ICASSP 2019}, May 2019, pp. 536--540.

\bibitem{mird}
E.~{Hadad}, F.~{Heese}, P.~{Vary}, and S.~{Gannot},
\newblock ``Multichannel audio database in various acoustic environments,''
\newblock {\em IWAENC 2014}, pp. 313--317, 2014.

\bibitem{timit}
J.~S. Garofolo, L.~F. Lamel, W.~M. Fisher, J.~G. Fiscus, D.~S. Pallett, and
  N.~L. Dahlgren,
\newblock ``{DARPA} {TIMIT} acoustic phonetic continuous speech corpus
  {CDROM},'' 1993.

\bibitem{Adam}
D.~Kingma and J.~Ba,
\newblock ``Adam: A method for stochastic optimization,''
\newblock in {\em ICLR 2015}, 2015.

\bibitem{tensorflow2015-whitepaper}
M.~Abadi, A.~Agarwal, P.~Barham, E.~Brevdo, Z.~Chen, C.~Citro, G.~S. Corrado,
  A.~Davis, J.~Dean, M.~Devin, S.~Ghemawat, I.~Goodfellow, A.~Harp, G.~Irving,
  M.~Isard, Y.Jia, R.~Jozefowicz, L.~Kaiser, M.~Kudlur, J.~Levenberg,
  D.~Man\'{e}, R.~Monga, S.~Moore, D.~Murray, C.~Olah, M.~Schuster, J.~Shlens,
  B.~Steiner, I.~Sutskever, K.~Talwar, P.~Tucker, V.~Vanhoucke, V.~Vasudevan,
  F.~Vi\'{e}gas, O.~Vinyals, P.~Warden, M.~Wattenberg, M.~Wicke, Y.~Yu, and
  X.~Zheng,
\newblock ``{TensorFlow}: Large-scale machine learning on heterogeneous
  systems,'' 2015,
\newblock Software available from tensorflow.org.

\bibitem{bsseval}
E.~Vincent, R.~Gribonval, and C.~Fevotte,
\newblock ``Performance measurement in blind audio source separation,''
\newblock {\em IEEE Transactions on Audio, Speech, and Language Processing},
  vol. 14, no. 4, pp. 1462--1469, July 2006.

\bibitem{Erdogan2015}
H.~{Erdogan}, J.~R. {Hershey}, S.~{Watanabe}, and J.~{Le Roux},
\newblock ``Phase-sensitive and recognition-boosted speech separation using
  deep recurrent neural networks,''
\newblock in {\em ICASSP 2015}, April 2015, pp. 708--712.

\end{thebibliography}
\end{document}